\documentclass[preprint,12pt]{elsarticle}

\usepackage{amssymb}

\journal{Nuclear Physics B}

\begin{document}

\begin{frontmatter}



\title{Two Particle Correlations at Forward Rapidity in STAR}
\author{Ermes Braidot, for the STAR Collaboration}
\ead{braidot@rcf.rhic.bnl.gov}
\address{Universiteit Utrecht,  Princetonplein 5, 3584CC Utrecht, The Netherlands}

\begin{abstract}
During the 2008 run RHIC provided high luminosity in both p+p and d+Au collisions at $\sqrt{s_{NN}}= 200$ GeV. Electromagnetic calorimeter acceptance in STAR was enhanced by the new Forward Meson Spectrometer (FMS), and is now almost contiguous from $-1<\eta<4$ over the full azimuth. This large acceptance provides sensitivity to the gluon density in the nucleus down to $x\approx 10^{-3}$, as expected for $2\rightarrow 2$ parton scattering. Measurements of the azimuthal correlation between a forward $\pi^{0}$ and an associated particle at large rapidity are sensitive to the low-x gluon density. Data exhibit the qualitative features expected from gluon saturation. A comparison to calculations using the Color Glass Condensate (CGC) model is presented.
\end{abstract}


\end{frontmatter}



\section{Introduction}

Two particle azimuthal correlations represent a powerful tool for characterizing the transitional region between dilute and saturated partonic systems. Inclusive particle production measurements from early RHIC runs show general agreement with pertubative QCD for p+p collisions at
$\sqrt{s}$=200 GeV over a broad range of rapidity.  Results from d+Au
collisions show Cronin type enhancements at midrapidity and strong
suppression for forward particle production \cite{PhysRevLett.97.152302, PhysRevLett.93.242303, PhysRevLett.91.072305}, suggestive of parton saturation,
although consistent with other explanations. The broad range of longitudinal momentum fraction $x$
that is averaged over in inclusive production is more selectively
probed in two particle correlations, especially when the
pseudorapidity ($\eta=-log(tan(\theta/2))$) of one of the two particles is
varied.

In pQCD at leading order, particle production in high energy hadronic
interactions results from the elastic scattering of two partons
($2\rightarrow 2$ scattering).  Each initial state parton carries a
fraction of the colliding hadron momentum, as given by universal
parton distribution functions. The scattered partons fragment to the
observed hadrons. Although complexities arise in calculating
two-particle correlations using pQCD, $2\rightarrow 2$ partonic scattering
leads to the expectation of events with back-to-back jets.  When
high-$p_T$ hadrons are used as jet surrogates, we expect the azimuthal
correlations of hadron pairs to show a peak at $\Delta\phi\sim\pi$
from back-to-back jets, and a peak at $\Delta\phi\sim0$ when
$\Delta\eta$ between the two particles is smaller than a typical jet
size.  Transverse momentum effects will broaden these peaks in the
azimuthal correlation function
 
When the parton density increases, the basic dynamics for the particle
production can change. Instead of elastic $2\rightarrow2$ scattering,
the particle production can proceed by the interaction of a probe parton
from the deuteron beam with multiple gluons from the Au beam.  At
sufficiently high parton densities, the transverse momentum from the
fragments of the probe parton may be compensated by several gluons
with lower $p_T$.  Two particle azimuthal correlations are expected to
show a broadening of the back-to-back peak (loss of correlation:
$2\rightarrow$ many processes) and eventually to disappear (monojet).

Different models try to describe hadron production from dense (nuclear) targets by including non-linear contributions. One approach \citep{2006PhLB..632..507Q} extends perturbative QCD factorization by adding contributions of quarks scattering coherently off multiple partons in the target. In the Color Glass Condensate (CGC) framework \citep{1983PhR...100....1G, 1986NuPhB.268..427M, 1994PhRvD..49.3352M, 2002PhRvL..89b2301D, 2004NuPhA.742..182I}, the hadronic wave-function is saturated as a consequence of gluon recombination. At very low values of the longitudinal momentum fraction $x$ of the probed gluons the occupation numbers become large. In both approaches the probe scatters coherently off the dense gluon field of the target, which recoils collectively, leading to a modification in $\Delta\varphi$. 

\section{Experimental Setup}

\begin{figure}
\includegraphics[width=0.5\textwidth, height=0.43\textwidth]{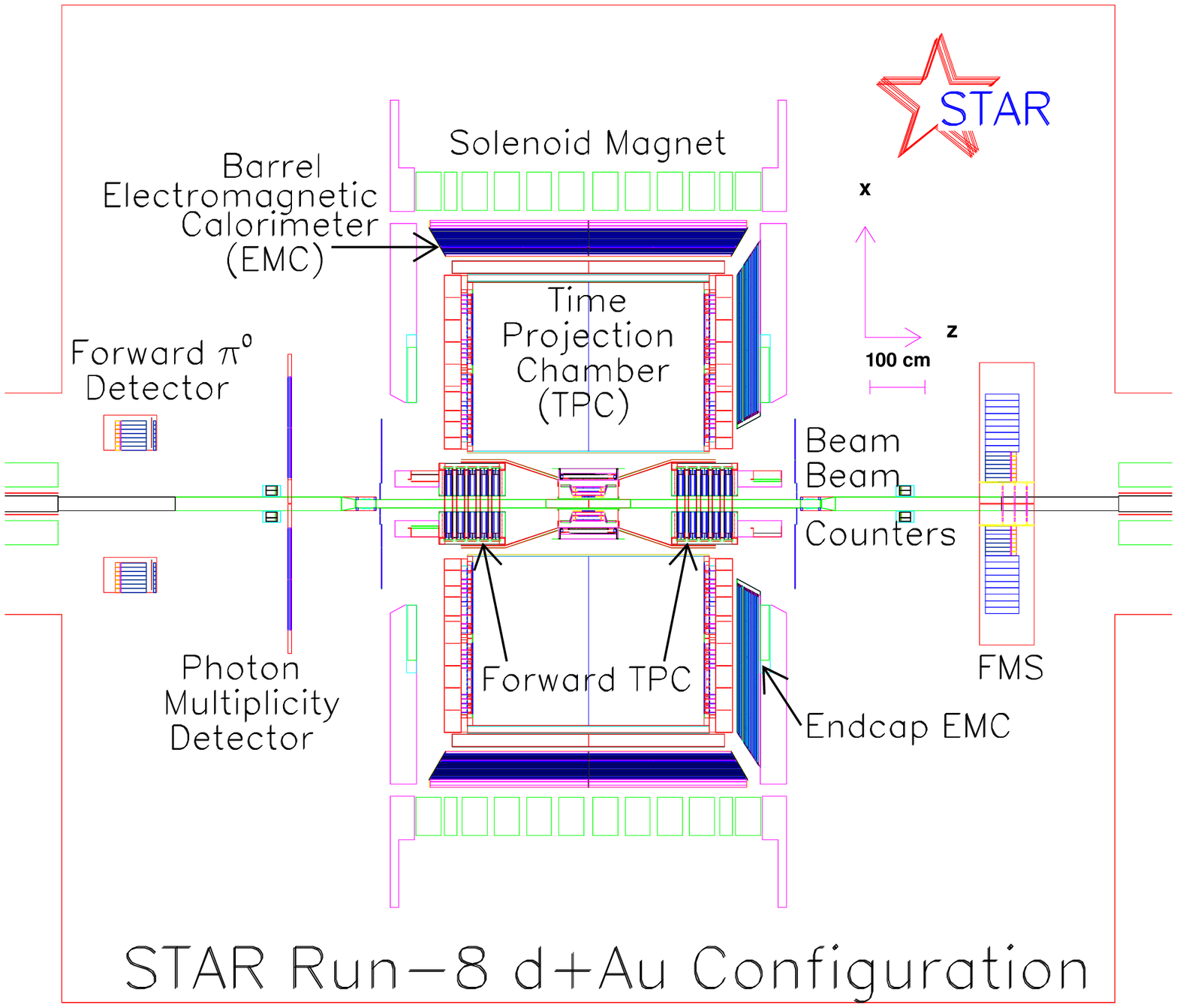}
\includegraphics[width=0.5\textwidth]{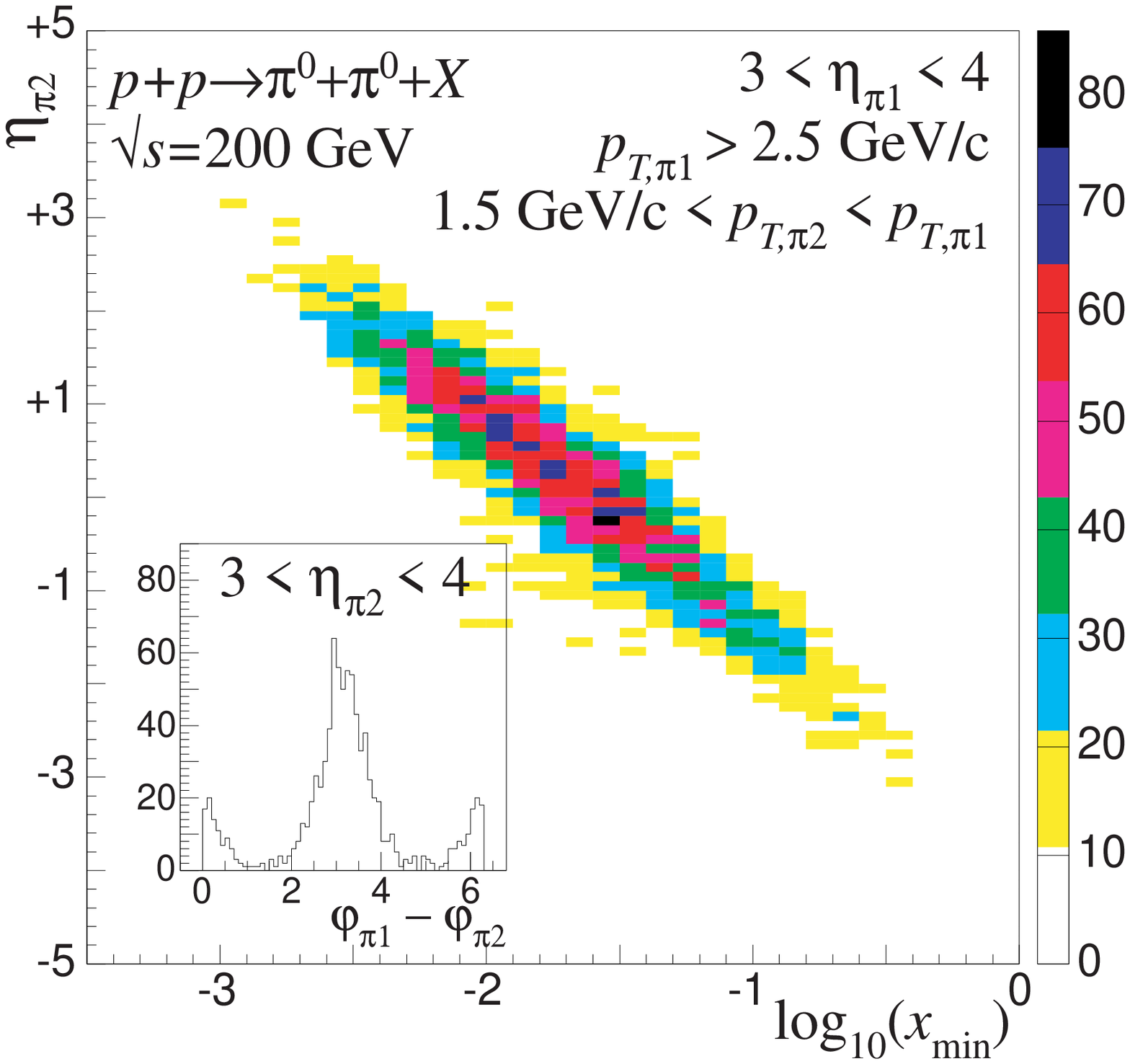}
\caption{On the left: schematic view of the STAR experimental hall. Gold beam is coming from the West (right in figure) side. On the right: PYTHIA \citep{Sjostrand:2001yu} simulation of di-pion production at large $\eta$ p+p collisions at $\sqrt{s}=200$ GeV. The $\eta$ of the associated particle is strongly correlated to the $x$ value of the soft parton probed in the partonic scattering. Figure from \citep{Bland:2005uu}.}\label{hall}
\end{figure}

The STAR detector at RHIC provides the suitable configuration for studying two particle azimuthal correlations. Electromagnetic calorimeters provide nearly
hermetic coverage for $2\pi$ in azimuth and  $-1.0 < \eta < 4.0$. The STAR calorimetry includes the Barrel ElectroMagnetic Calorimeter (BEMC: $|\eta|<1.0$), the End-cap ElectroMagnetic Calorimeter (EEMC: $1.0<\eta<2.0$) and the newly installed Forward Meson Spectrometer (FMS: $2.5<\eta<4.0$). These, together with the STAR Time Projection Chamber (TPC), allow measurements of correlations of different particles over a broad $\Delta\eta\times\Delta\varphi$ range.

Signatures of possible saturation effects can be investigated by comparing azimuthal correlations in p+p and d+Au interactions. The first case is used as a reference: the colliding systems are dilute and no saturation effects are expected at these energies. In d+Au, on the contrary, the dense nuclear matter is probed and azimuthal correlations are expected to broaden when the saturation scale is reached. No large final state rescattering effects (as in Au-Au collisions) are expected in d+Au. The measurement of di-hadron
azimuthal correlation functions involving a forward neutral pion, and
variable $\eta$ for the second hadron, are expected to reveal
signatures of saturation when the gluons are at sufficiently low $x$.
The STAR trigger systems allow us to select events with a leading $\pi^{0}$ within the FMS acceptance.
In a pQCD description, neutral pions are produced in the
forward direction by the interaction between a large-$x$ parton
(most likely a valence quark) from the beam heading towards the FMS
(either p or d) with a low-$x$ gluon from the other beam (either p or
Au). The correlation probability is then calculated by detecting a second particle in any of the STAR sub-detectors. Since the saturation scale depends both on the momentum fraction $x$ of the probed gluon and on the scale $Q^2$, one can approach the saturation region by requiring an associated particle with increasing pseudo-rapidity, or by selecting events characterized by a lower $p_{T}$ respectively. Additional insights can be achieved by selecting correlations based on the centrality of the collision, since a larger effect is expected when the denser part of the nucleus is probed.   

\section{Results}

A plan of azimuthal correlation measurements \citep{Bland:2005uu} has been systematically pursued at STAR. The objective has been to determine if the boundaries of the saturation region are accessible at RHIC energies \citep{2010PhLB..687..174A} and establish the effects on particle production. The $p_{T}$ dependences of high density nuclear effects have been studied using prototype forward calorimeters (Forward Pion Detector, FPD/FPD++, \citep{PhysRevLett.97.152302}) and recently the FMS \citep{Braidot2009603c}.  Comparison of $\Delta\varphi$ between a forward $\pi^{0}$ and a mid-rapidity hadron in p+p and d+Au shows a significant broadening in the back-to-back peak in d+Au. Such effect appears to be stronger as the $p_{T}$ of the particles decreases, as expected from saturation models.
The $x$ dependence is accessed by comparing the effects on $\Delta\varphi$ while varying the pseudo-rapidity of the associated particle. Correlations between two forward pions \citep{Braidot:2010zh} show indeed a much stronger broadening than in forward + mid-rapidity correlations. Measurements in the intermediate $\eta$ region (using the STAR EEMC) are ongoing. The centrality dependence of the effect was studied for all $\eta$ and $p_{T}$ regimes. In this analysis, the impact parameter $b$ of the collision is evaluated in relation to the multiplicity, measured as the sum of charges ($\sum{Q_{BBC}}$) in the event by the east side of the STAR Beam-Beam Counter (BBC), facing the Au beam. The multiplicity classes for peripheral ($0<\sum{Q_{BBC}}<500$) and central d+Au collisions ($2000<\sum{Q_{BBC}<4000}$) correspond to $\langle b \rangle=(6.8\pm1.7)$ fm and $\langle b \rangle=(2.7\pm 1.3)$ fm respectively, as indicated from a HIJING 1.383 \citep{PhysRevD.44.3501} simulation.

\def\imagetop#1{\vtop{\null\hbox{#1}}}
\begin{figure}
\begin{tabular}{c c c}
  \imagetop{\includegraphics[height=0.3\textwidth]{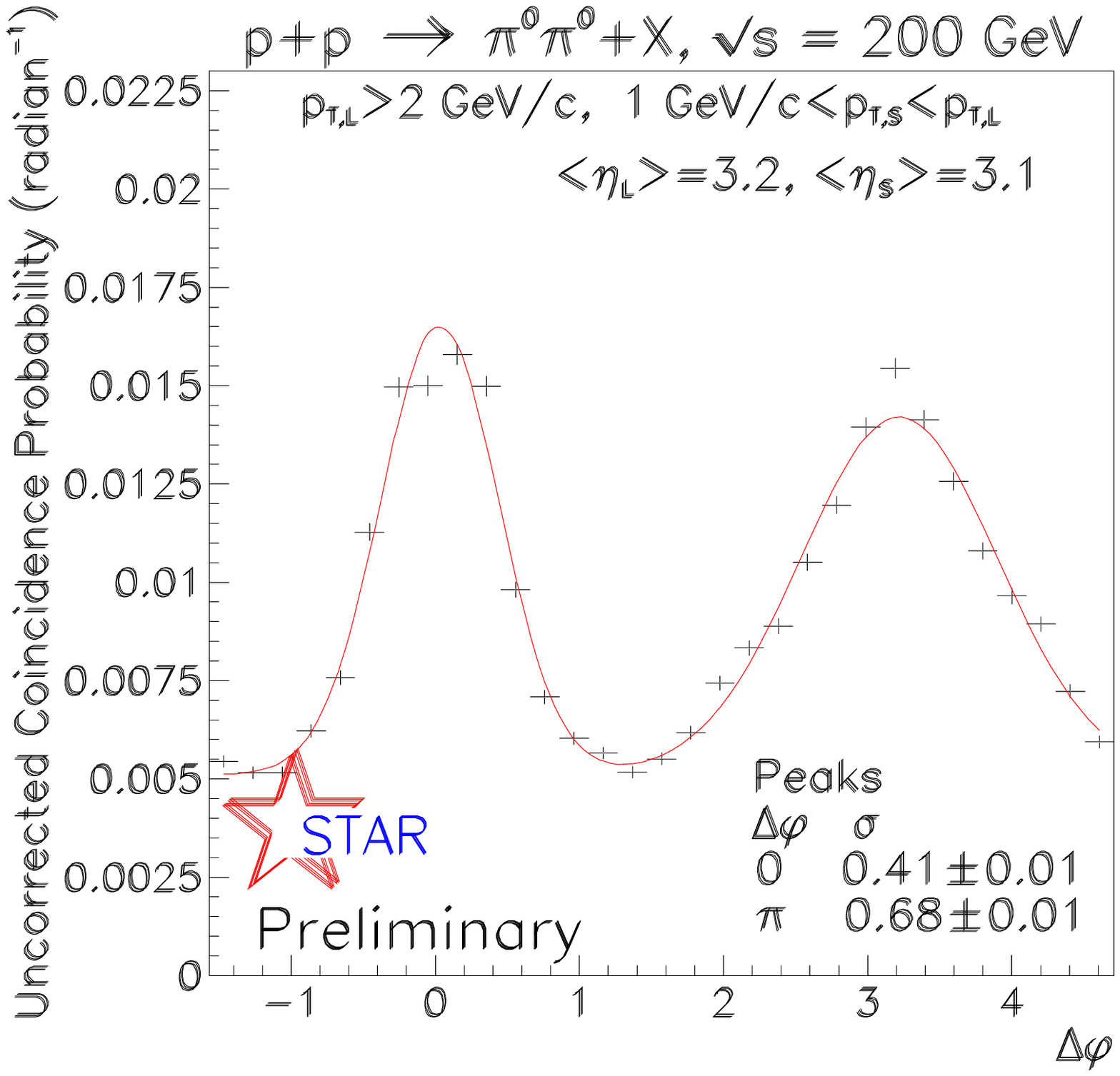}}
  \imagetop{\includegraphics[height=0.3\textwidth]{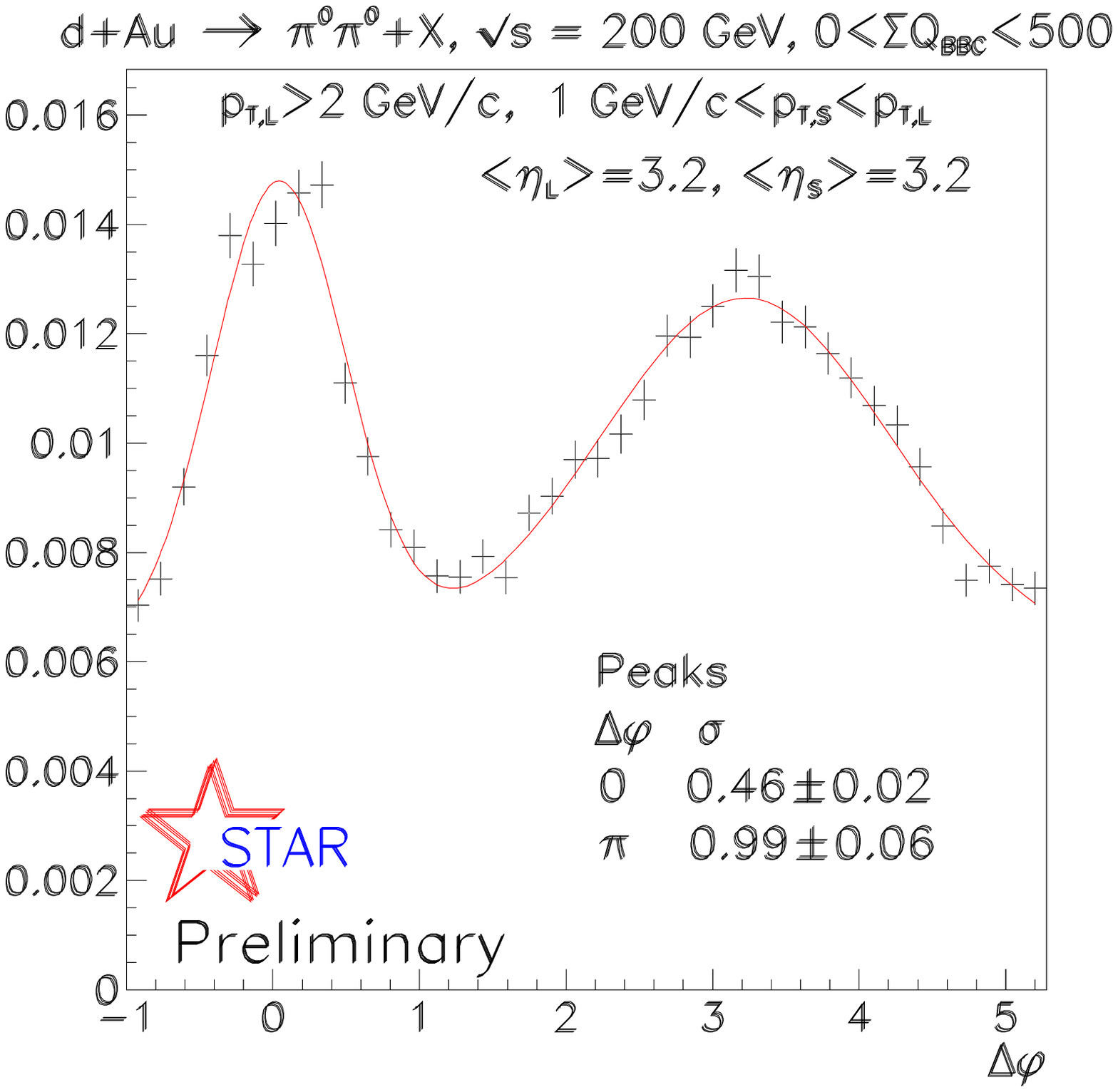}}
\imagetop{\includegraphics[height=0.285\textwidth]{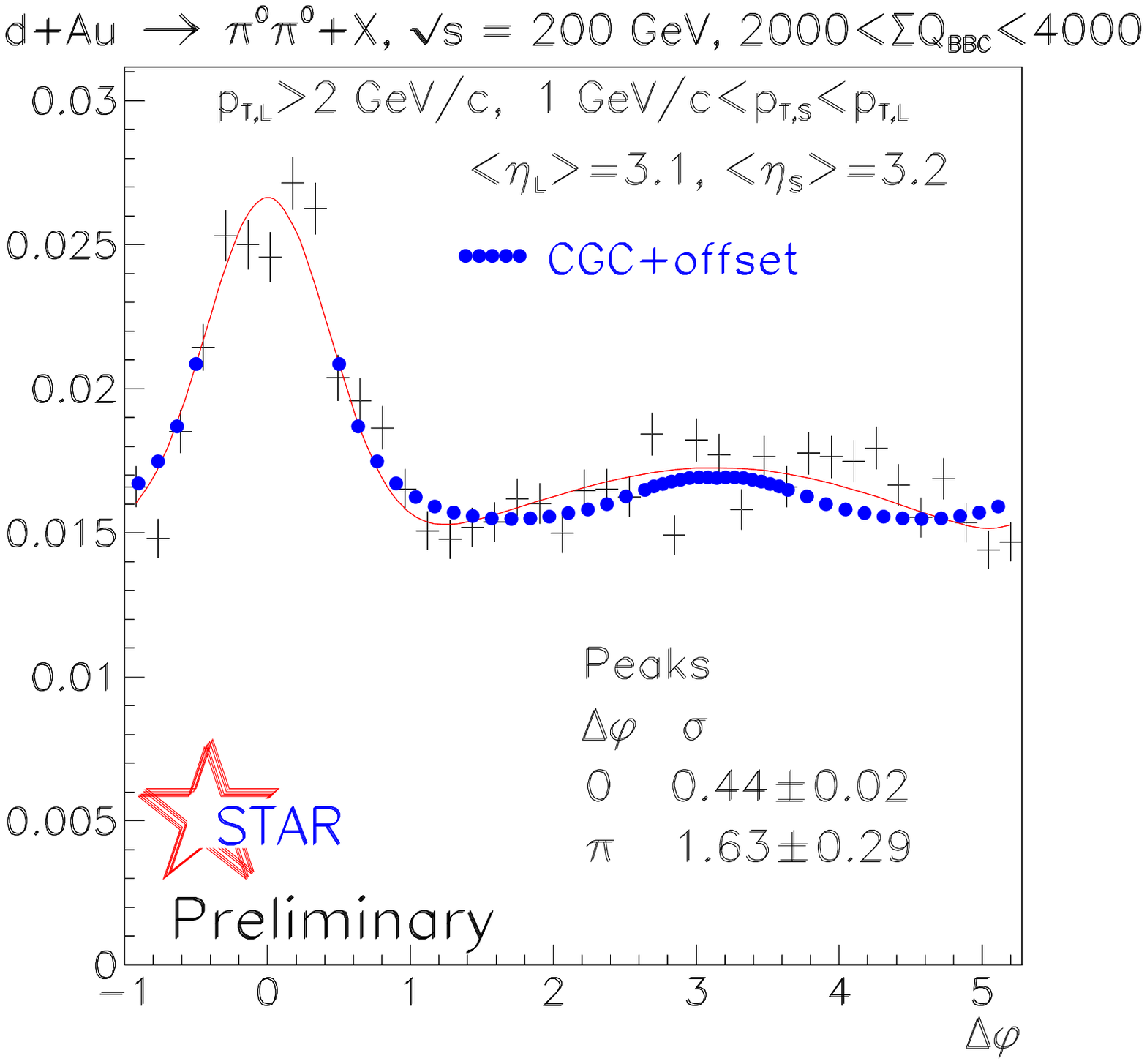}}
\end{tabular}
\caption[]{Uncorrected coincidence probability versus azimuthal angle difference between two forward neutral pions in p+p collisions (left) compared to peripheral (center) and central d+Au collisions (right). Data are shown with statistical errors and fit with a constant plus two Gaussian functions (in red). CGC expectations \citep{Marquet200741} have  been superimposed (in blue) to data for central d+Au collisions.}
\label{central}
\end{figure}

Figure \ref{central} shows the (efficiency uncorrected) probability to find an associated pion given a trigger $\pi^{0}$, both in the forward (FMS) region. Neutral pions are reconstructed from pairs of photon clusters within the FMS fiducial volume. Pion candidates are required to have an invariant mass in the interval $0.05<M_{\gamma\gamma}<0.25$ GeV/c$^2$. The pair with the largest $p_{T}$ is selected as the leading (trigger) pion and its azimuthal coordinate is compared inclusively with those of all other (associated) pion candidates. Trigger particles are selected with a transverse momentum of $p^{(trg)}_{T}>2.0$ GeV/c while associated particles are required to have $p^{(trg)}_{T}>p^{(assc)}_{T}>1.0$ GeV/c. In Figure \ref{central} we compare $\Delta\varphi$ for p+p interactions with peripheral and central d+Au collisions.
All the distributions present two signal components, surmounting a constant background representing the underlying event contribution (larger in d+Au). The peak centered at $\Delta\varphi=0$ (\emph{near-side peak}) represents the contribution from pairs of pions belonging to the same jet. It is not expected to be affected by saturation effects, therefore it is a useful tool to check the effective amount of broadening in the \emph{away-side peak}. This second peak, centered at $\Delta\varphi=\pi$, represents the back-to-back contribution to the coincidence probability which is expected to disappear in going from p+p to d+Au if saturation sets in. Data are fit with a constant plus two Gaussians centered at $\Delta\varphi=0$ and $\Delta\varphi=\pi$. Data show that the width of the near-side peak remains nearly unchanged from p+p to d+Au, and particularly from peripheral to central d+Au collisions. The away-side peak, instead, presents strong differences between peripheral and central d+Au collisions. Peripheral d+Au collisions show back-to-back contribution like in p+p, even though apparently smaller in height
relative to the near-side peak than in p+p, but broadened. Central d+Au collisions show a substantially reduced away side peak that is significantly broadened.  

The rightmost panel of Figure \ref{central} shows also a comparison (in blue) with  theoretical expectations using the CGC framework. Inclusive particle production in interactions between a dilute system (deuteron) and a saturated target (Gold) has been calculated using a fixed saturation scale $Q_{S}$ \citep{2010PhLB..687..174A} and, consequently, used to compute di-hadron correlations \citep{Marquet200741}. Calculations consider valence quarks in the deuteron scattering off low-$x$ gluons in the nucleus with impact parameter b = 0. 
CGC calculations show qualitative consistency with data in their expectations of a strong suppression of the away-side peak in central d+Au collisions. 

\section{Systematics}

\begin{figure}
\begin{tabular}{c c}
  \imagetop{\includegraphics[height=0.5\textwidth, width=0.4\textwidth]{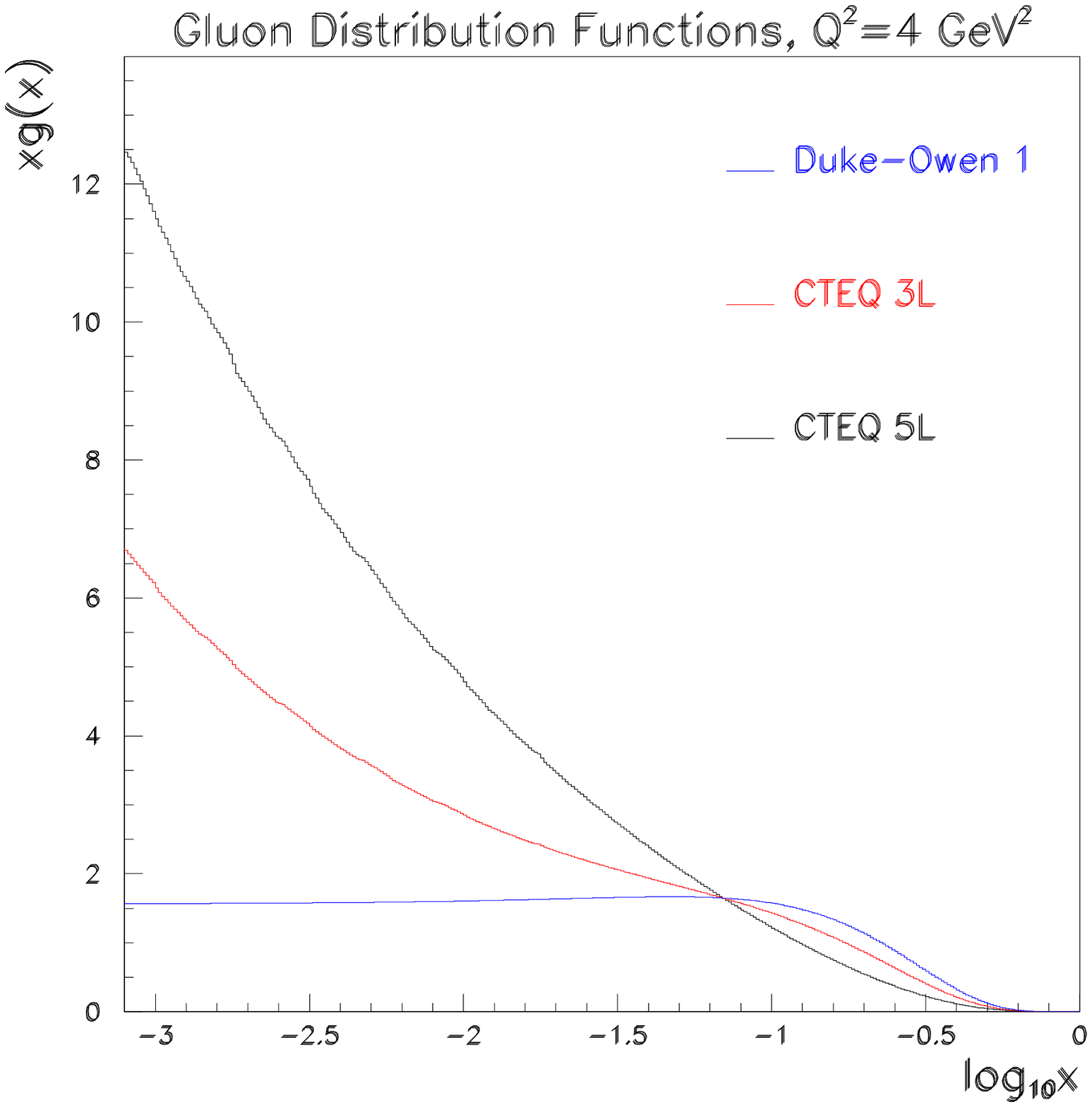}}
  \imagetop{\includegraphics[height=0.5\textwidth, width=0.57\textwidth]{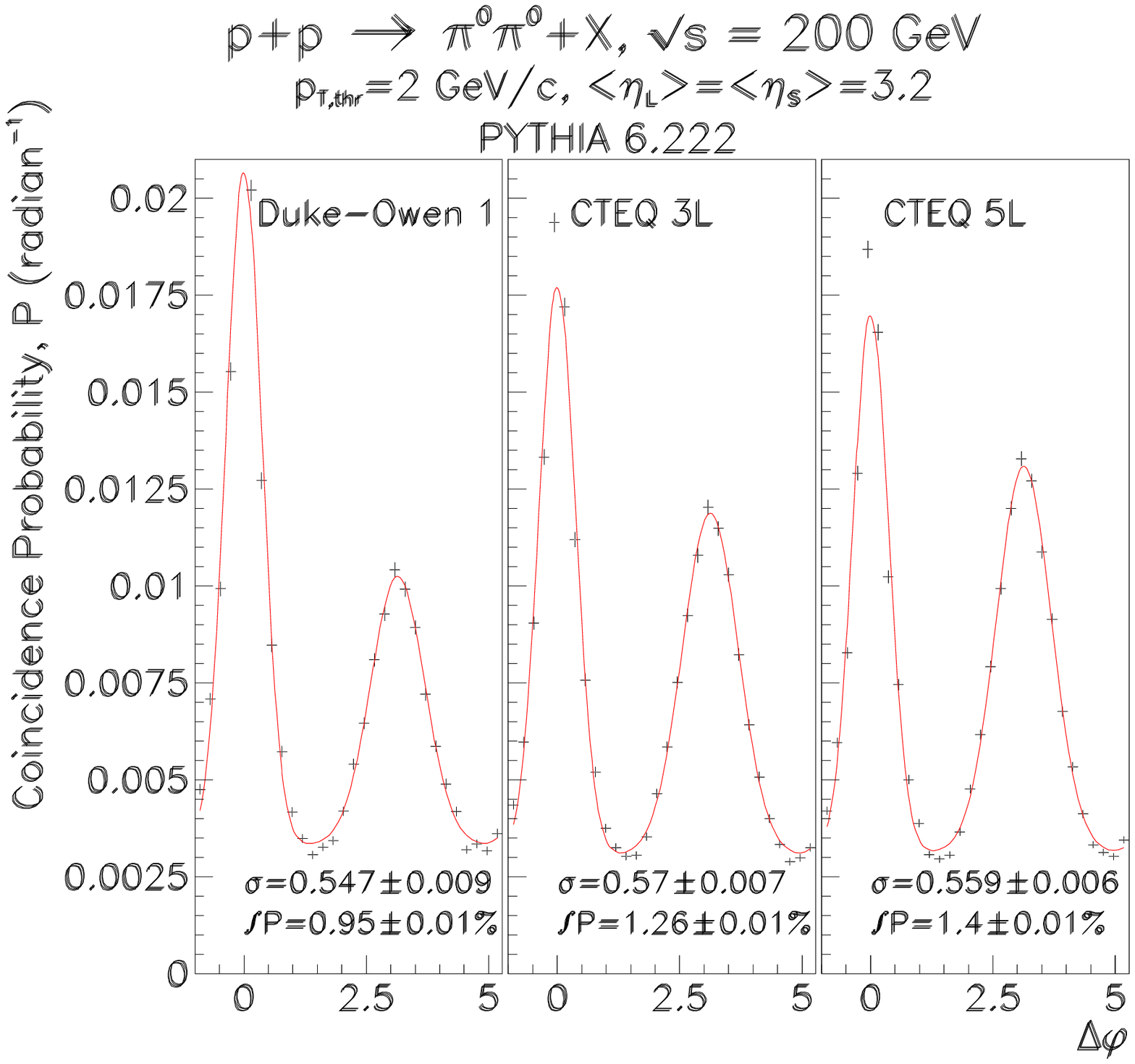}}
\end{tabular}
\caption[]{Comparison between parton density functions and their effect on azimuthal correlations. Left: comparison of density distributions of gluons using Duke-Owen (blue), CTEQ3L (red) and CTEQ5L  (black) PDF sets. Right: uncorrected forward di-pion correlations in p+p as simulated in PYTHIA 6.222, using the three different PDF sets. }
\label{pdf}
\end{figure}

Further studies have been performed on this analysis with the goal of evaluating the expectations of ``conventional'' calculations of azimuthal correlations, using PYTHIA 6.222 and HIJING 1.383. Forward di-hadron correlations are expected to be sensitive to gluon density, initial-state showers and intrinsic transverse momentum $k_{T}$ of the partons. HIJING models d+Au collisions by using a
version of PYTHIA, and a Glauber model for the nuclear effects. The latest available HIJING versions present only a limited choice of parton distributions functions (PDF), all of which predate HERA discovery of the rapid growth of the gluon distribution at low-$x$. A systematic study of the dependence of forward di-pion
azimuthal correlations on the PDF has been made using PYTHIA 6.222. Figure \ref{pdf} shows the comparison between three different sets of parton distributions: the Duke-Owen set \citep{PhysRevD.26.1600}, which predates HERA data \citep{Adloff:1999kg, Cheka} and therefore does not include a rise in the gluon density at low-$x$, and the two sets CTEQ3L and CTEQ5L \citep{Lai:1999wy}, the latter with a steeper rise of the gluon distribution at low-$x$. The right panel in Figure \ref{pdf} shows the azimuthal correlations expected from these three PDF sets in p+p interactions. Data appear to be consistent in both peak areas and widths with a set of PDF that presents a rapid rise of the gluon density at low-$x$. 

\begin{figure}
  \imagetop{\includegraphics[width=1.0\textwidth]{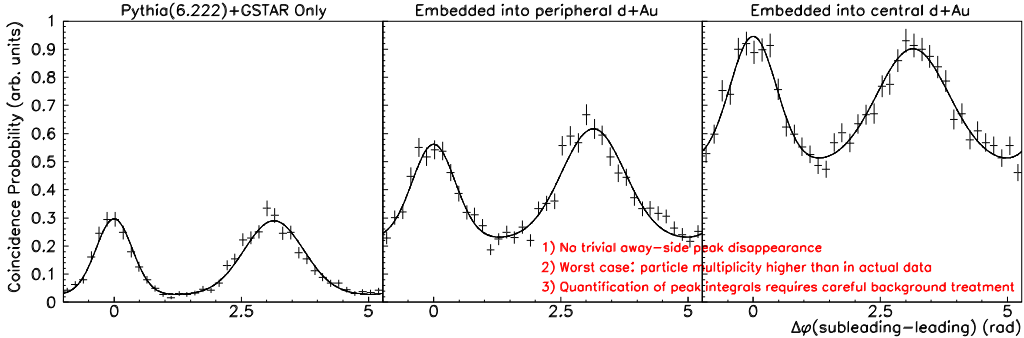}}
\caption[]{Uncorrected azimuthal correlations between two forward neutral pions as simulated in p+p by PYTHIA 6.222 (left). On the central and rightmost panel: PYTHIA events have been embedded into minimum bias d+Au data to emulate the effect of the larger multiplicity in d+Au. Each probability is scaled by a common arbitrary factor.}
\label{embedding}
\end{figure}

The effect on azimuthal correlations of the larger multiplicity in d+Au compared to p+p events has been studied. For this purpose, simulated di-pion PYTHIA events have been embedded into minimum bias d+Au data. Embedding has been preferred to pure HIJING simulations for reasons discussed before. Figure \ref{embedding} shows the results of these studies. Azimuthal correlations between two forward pions are simulated using PYTHIA only and compared to PYTHIA events embedded into minimum bias peripheral and central d+Au interactions. The effect of the additional multiplicity in d+Au affects particularly the reconstruction of the associated $\pi^{0}$, causing an higher combinatorial background which reflects in the higher background level seen in d+Au in simulations as well as in data. The features of the near-side peak are reproduced in both p+p and d+Au in this embedding simulation. On the other side, the large suppression of the away-side peak noted in data, in particular in central d+Au collisions, is not reproduced in this simulation. We conclude that the away-side peak suppression is not the result of the multiplicity increase in central d+Au
collisions relative to p+p collisions.

\begin{figure}
\begin{tabular}{c c c}
  \imagetop{\includegraphics[height=0.31\textwidth]{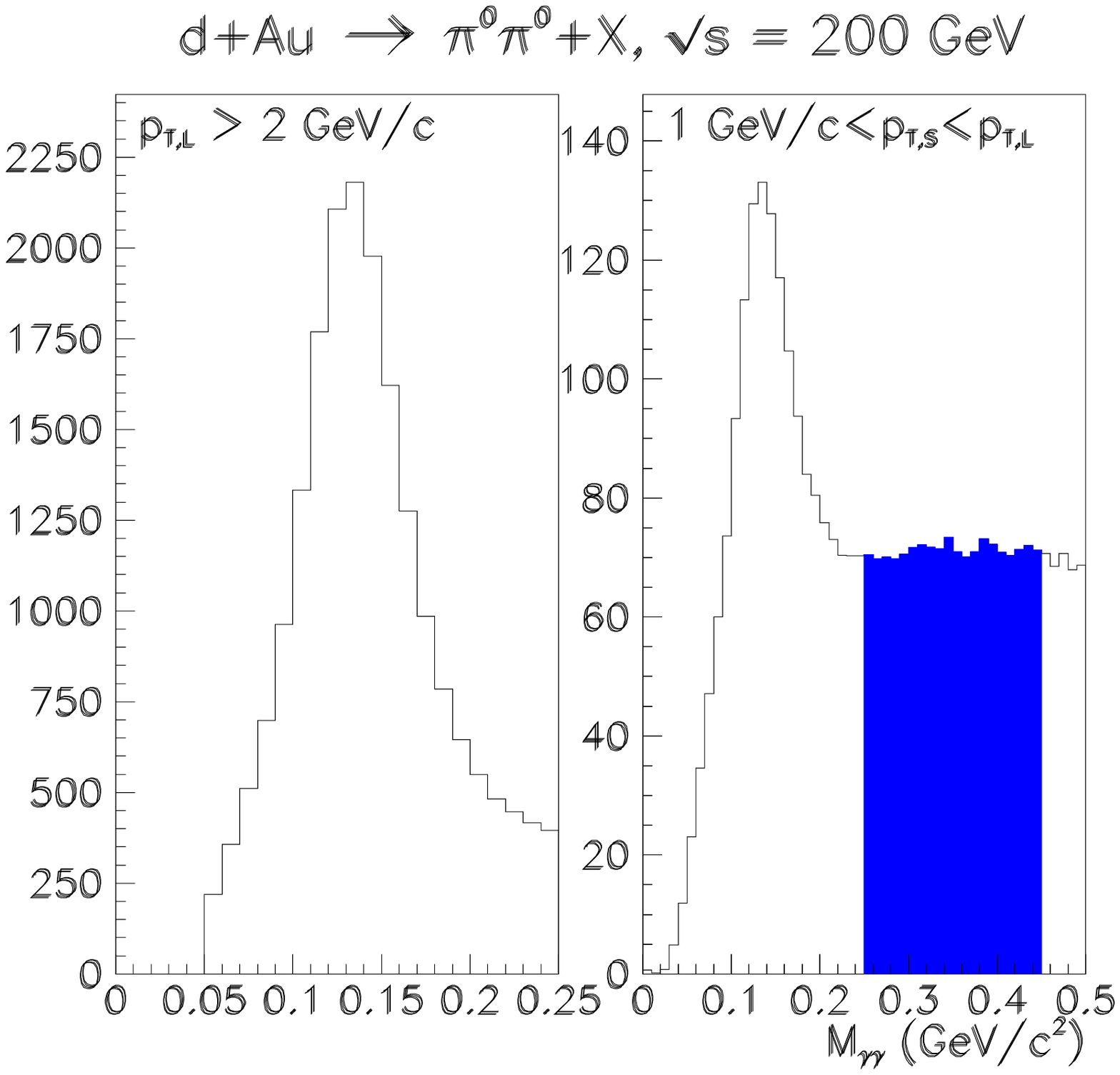}}
  \imagetop{\includegraphics[height=0.31\textwidth]{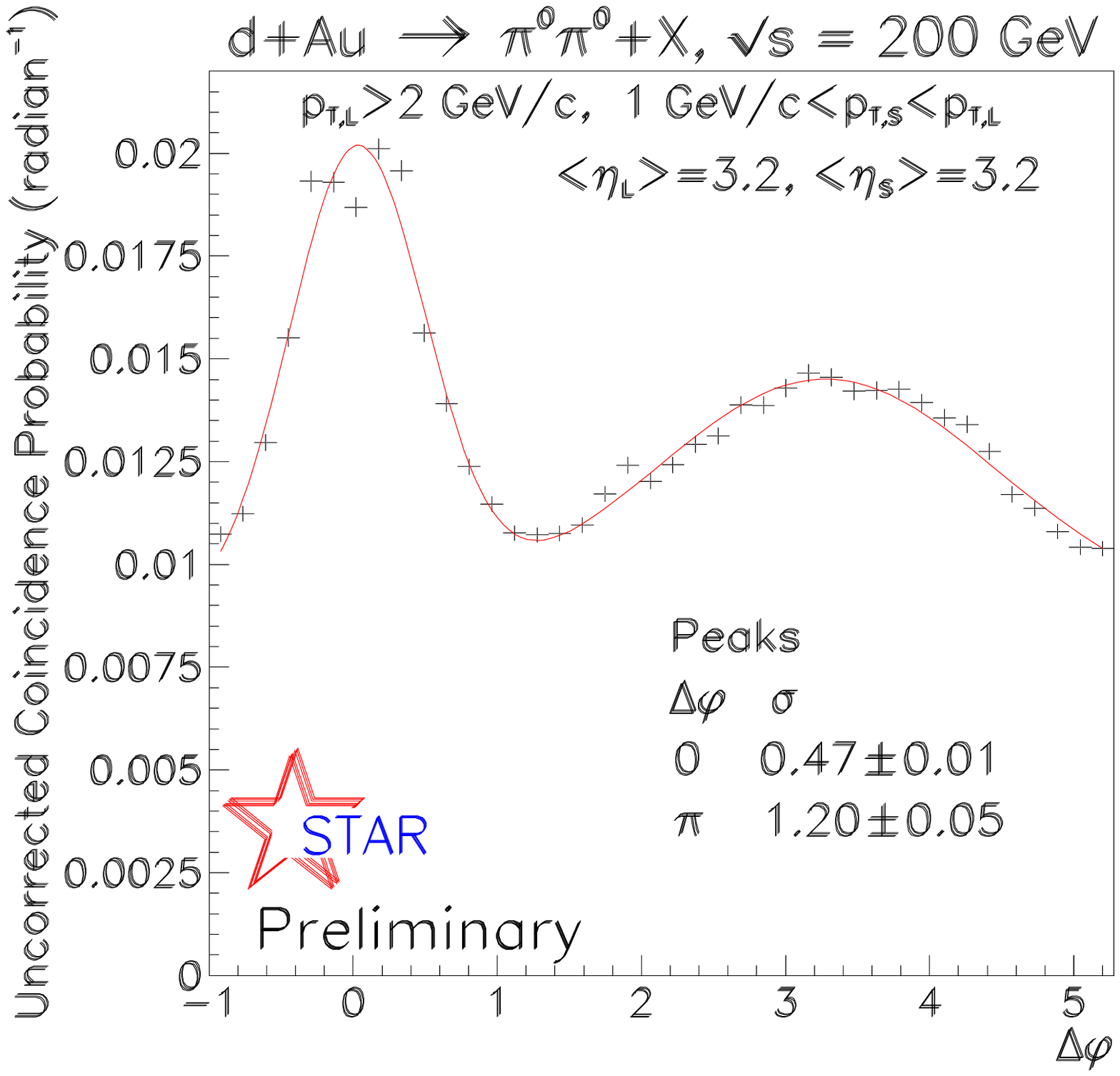}}
  \imagetop{\includegraphics[height=0.31\textwidth]{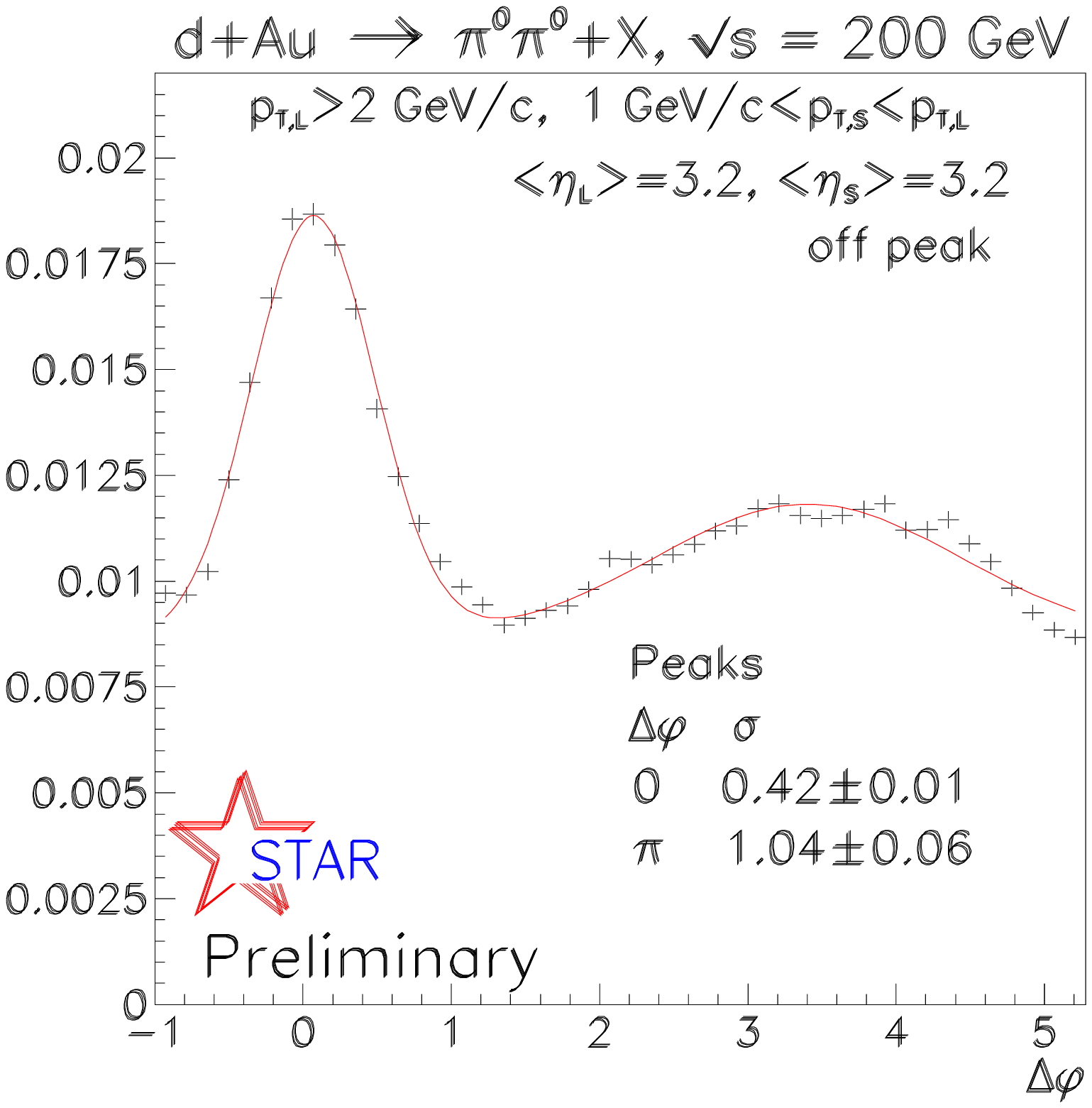}}
\end{tabular}
\caption[]{Off-peak analysis. On the left panel invariant mass spectra for leading and sub-leading pion candidates are shown. The off-mass selection for sub-leading pions is indicated in blue. Following, uncorrected azimuthal correlations between two forward pion candidates (center) and between a forward pion candidate and a pair of clusters with invariant mass in the range $0.25<M_{\gamma\gamma}<0.45$ GeV/c$^{2}$ (right).}
\label{offpeak}
\end{figure}

As a further test, the impact of the combinatorial background on the shape of the azimuthal correlations has been studied. To do so, a pair of clusters with an invariant mass in the range $0.25<M_{\gamma\gamma}<0.45$ GeV/c$^{2}$, relatively far from the nominal value of the $\pi^{0}$ mass, has been used as associated particle. Figure \ref{offpeak} shows the azimuthal correlation between a leading $\pi^{0}$ and such associated pairs. With this mass selection, $\Delta\varphi$ appears to be quantitatively consistent, in the widths of both peaks, with the ``on-peak'' correlation. Such small impact is expected for jet-like correlations where the combinatorial background is produced between pairs in the same jet. 

\section{Conclusions}

Two particle correlations are a more sensitive probe of
the transition region between dilute and saturated systems than
inclusive particle production. Thanks to a rich d+Au RHIC run in 2008, STAR is pursuing its objective of mapping the boundaries of the saturation region. Two particle azimuthal correlations have been systematically studied as a function of $p_{T}$, $\eta$ and multiplicity. When two hadrons are reconstructed in the forward region, where the lowest $x$ in the gluon distribution is probed, comparisons with simulation indicates that p+p data are sensitive to the low-$x$ gluon density. Measurements over p+p and d+Au data show a strong suppression of the away-side peak in central d+Au collision compared to p+p, qualitatively consistent with CGC calculations. Systematic studies show that such suppression is not expected to arise from additional multiplicity in d+Au and it is not significantly affected by the presence of combinatorial background in the invariant mass di-photon spectrum.





\bibliographystyle{elsarticle-num}
\bibliography{bibliography.bib}







\end{document}